\documentclass[12pt]{article}

\usepackage{cite}
\usepackage{epsf}

\title{
\begin{flushright}
{\normalsize Yaroslavl State University\\
   Preprint YARU-HE-92/01 \\[5mm]
   hep-ph/0007149}\\[5mm]
\end{flushright}
Compton-like interaction of massive neutrinos with virtual photons}

\author{A.V.~Kuznetsov and N.V.~Mikheev\\
{\small\it Department of Physics, Yaroslavl State University,}\\
{\small\it 150000 Yaroslavl, Russian Federation.}}

\date{}

\begin{document}

\maketitle

\begin{abstract}

The amplitude of a Compton-like process $\nu_{i}\gamma^{*}\rightarrow 
\nu_{j}\gamma^{*}$ with virtual photons is calculated in the standard 
GWS theory with
lepton mixing. The contribution of this process to the high energy
neutrino scattering on the nucleus with single photon radiation $\nu N 
\rightarrow \nu N\gamma $
is discussed. The bremsspectrum and the total cross-section are
estimated in the leading $\log $ approximation.

\end{abstract}

\vglue 5mm

\begin{center}
{\it Published: Physics Letters B 299 (1993) 367-369} 
\end{center}

\newpage

Electromagnetic properties of neutrinos as higher-order
effects of weak interactions are of considerable interest and have
a long history of discussions. Their possible manifestations are
the Compton-like photon-neutrino scattering $\nu \gamma \rightarrow 
\nu \gamma $ and crossed
reactions. However, as was shown by Gell-Mann~\cite{GellM} in the local
four-fermion $(V-A)$ theory of weak interaction, the amplitude of
such a process becomes zero when massless neutrinos and real
photons are considered. It can be understood if $\nu \bar{\nu } 
\rightarrow \gamma \gamma $ process is
analyzed in the center-of-mass system. In this case the total
angular momentum of the neutrino-antineutrino pair can be equal
only to a unit while the wave-function of the two on-shell photon
system does not have such a state~\cite{Land}. The amplitude becomes
nonzero if any of the Gell-Mann theorem conditions is broken. It
may be non-locality of neutrino interaction, non-zero neutrino
mass or off-shell photons. The initial calculation of 
$\gamma^{*}\gamma \rightarrow \nu \bar{\nu }$
amplitude with one off-shell photon and $m_{\nu }=0$ was made by 
Rosenberg~\cite{Rose}
also in the frame of four-fermion theory. In that paper the
$\gamma \rightarrow \nu \bar{\nu }$ process in the coulombian field 
of the nucleus was considered
as an additional mechanism of the energy loss by stars. The
leading local term for $\nu \gamma^{*}\rightarrow \nu \gamma^{*}$ 
with two virtual photons and $m_{\nu }=0$
was found~\cite{Cung} in the gauge theory of weak interaction. An exact
expression for the amplitude of the process $\gamma \gamma \rightarrow 
\nu \bar{\nu }$ with massive
neutrinos and on-shell photons was obtained for the first time in
ref.~\cite{Crew} (see also ref.~\cite{Dode}) to estimate the star energy 
loss due to this reaction.

Here we find the most general expression for the amplitude of
$\nu_{i}\gamma^{*}\rightarrow \nu_{j}\gamma^{*}$ process 
(in general $i\neq j)$ in the standard model of
electroweak interaction embracing all the cases of real and
virtual photons, massive and massless neutrinos and taking into
account the mixing in the lepton sector. We consider the
approximation $(pq_{1,2})\ll m^{2}_{W}$ where $q_{1,2}$ and $p$ 
are the four-momenta
of photons and neutrino, respectively. The diagrams that give the
main contribution to the process in this limit are shown in figs. 1
and 2. Then the amplitude can be written in the form

\begin{eqnarray}
M & = & {\alpha \over \pi } \;{G_{F}\over \sqrt{2}} \;j^{(\nu )}_{\rho }
\left (\sum_{l} V^{*}_{il} V_{jl} R^{(l)}_{\rho } \; + \; \delta_{ij} 
\sum_{f} T_{3f} Q^{2}_{f} R^{(f)}_{\rho } \right ), 
\label{eq:1}\\
R^{(f)}_{\rho } & = & 4i \;\lbrace {1\over 2} (F_{1} \tilde F_2) 
(q_{2}-q_{1})_{\rho } \, 
A(m_{f},q_{1},q_{2}) 
\nonumber\\
& - & (\tilde F_{2} F_{1} q_{1})_{\rho } \, 
B(m_{f},q_{1},q_{2}) 
 +  (\tilde F_{1} F_{2} q_{2})_{\rho } \, B(m_{f},q_{2},q_{1}) \rbrace, 
\label{eq:2}\\
(F_{1}\tilde{F}_{2}) & = & F_{1\mu \nu } \tilde{F}_{2\nu \mu } , \qquad
(\tilde{F}_{2}F_{1}q)_{\rho }=\tilde{F}_{2\rho \mu }F_{1\mu \nu }q_{1\nu } ,
\nonumber\\
F_{\mu \nu } & = & q_{\mu }e_{\nu }-q_{\nu }e_{\mu } , \qquad
\tilde{F}_{\mu \nu }={1\over 2} \epsilon_{\mu \nu \alpha \beta } 
F_{\alpha \beta }, 
\label{eq:3}\\
&&A(m,q_{1},q_{2})  =  \int^1_0 x dx \int^{1-x}_0 y dy \;{1\over a} , 
\nonumber\\
&&B(m,q_{1},q_{2})  =  \int^1_0 x dx \int^{1-x}_0 dy \;{{1-x-y} \over a} , 
\label{eq:4}\\
a & = & m^{2}+2(q_{1}q_{2}) \;xy-q^{2}_{1} \; x(1-x)-q^{2}_{2} \; y(1-y).
\label{eq:5}
\end{eqnarray}

Here $j^{(\nu )}_{\rho } = \bar \nu_{j}(p_{2})\gamma_{\rho }(1-\gamma_{5})
\nu_{i}(p_{1})$ is the neutrino $(V-A)$ current,
$e_{\mu }$ is the photon polarization four-vector, $V_{il}$ is the mixing
matrix of Kobayashi-Maskawa type in the lepton sector. The first
term in (1) is related to the $W$-contribution (fig.1) and needs the
summation over all charged leptons. The second term in (1) comes
from the $Z$-contribution (fig.2) where the sum runs over all charged
fermions (both leptons and quarks). Here $eQ$ is the electric charge
of the fermion and $T_{3f}$ is the third component of weak isospin. Let
us note that the first term in eq.(2) is reduced to the divergence
of the neutrino current and so it will be proportional to the
neutrino mass. The obtained amplitude is explicitly gauge
invariant since it is expressed in terms of electromagnetic
tensors of the photons (3). In some particular cases it can be
reduced to the well-known results of refs.~\cite{Rose,Cung,Crew}. 
For example, if we assume
both photons to be virtual $(q^{2}_{1,2}\neq 0)$ and neutrino to be massless,
the amplitude (1), (2) can be transformed to the one given by Cung and
Yoshimura~\cite{Cung}. In our opinion, their amplitude contains the
artificial dependence on the neutrino momenta. It is obvious,
however, that in the considered approximation (that is in fact the
local limit of weak interaction) the amplitude of the $\nu \gamma^{*}
\rightarrow \nu \gamma^{*}$ process
can manifestly depend on the photon momenta only.

Our general result (1) allows us also to find the first terms in
the expansion for the amplitudes of the neutrino radiative decay
$\nu_{i}\rightarrow \nu_{j}\gamma $ and non-radiative transition 
$\nu_{i}\rightarrow \nu_{j}$ in the external
electromagnetic field of arbitrary configuration. For this purpose
it is sufficient to replace the electromagnetic tensor of either
one or both photons by the external field tensor.

We shall illustrate the result (1) considering the high
energy neutrino scattering process in the Coulombian field of a
nucleus with one photon radiation. Previously~\cite{Rose,Crew,Dode} 
only astrophysical effects of the process $\nu \gamma \rightarrow \nu 
\gamma $ were studied. Our aim
is to examine the possibility of the detection of the process 
in the laboratory. Really, it could be observed as bremsstrahlung when
the neutrino is scattered by a nucleus without its break-up,

\begin{equation}
\nu + nucleus \rightarrow \nu + \gamma + nucleus. 
\label{eq:6}
\end{equation}

\noindent The reaction amplitude can be obtained from (1), (2) taking one of
the photons (e.g. $\gamma_{2})$ to be real. In this case one has 
$F_{2\mu \nu }q_{2\nu }=0.$
To get over the technical difficulties we shall regard $m_{\nu }=0$ and
neglect the lepton mixing. Then the amplitude will be defined by
the second term in (2). Inserting $(Ze/q^{2}_{1})J_{\mu }$ instead of 
$e_{1\mu }$, where
$J_{\mu }$ and $Ze$ are the electromagnetic current and the charge of the
nucleus, $q_{1\mu }$ and $e_{1\mu }$ are the momentum and the polarization 
vector of the virtual photon, one obtains

\begin{equation}
M = 4i{Ze\alpha \over \pi } {G_{F}\over \sqrt{2}} 
\epsilon_{\rho \mu \alpha \beta }j^{(\nu )}_{\rho } 
J_{\mu } q_{2\alpha } e_{2\beta } 
\left ( B(m_{\ell},q_{1},q_{2}) + 
\sum_{f} T_{3f} Q^{2}_{f} B(m_{f},q_{1},q_{2}) \right ). 
\label{eq:7}
\end{equation}

\noindent Here $m_{\ell}$ is the mass of the charged lepton which is the 
partner of
the neutrino taking part in the reaction. Let us examine the case
of small transmitted momenta when the nucleus is still nearly
motionless. The momentum modulo $|\vec{q}_{1}|$ is restricted then by the
value of $q_{m}$ which can be estimated as the inverted nucleus radius
$q_{m}\simeq 1/r \simeq 300$ MeV. One can easily see from eq.(4) 
that at high
energies of the neutrino all the charged fermions contribute to the
amplitude (7) except $t$-quark (we still presume $(pq_{1})\ll m^{2}_{W}
< m^{2}_{t})$. In
the leading $\log $ approximation we get the following expression for
the spectrum of radiated photons:

\begin{equation}
d\sigma = {\alpha \over 54\pi } {\left ({Z \alpha } \over \pi \right )}^{2} 
{G^{2}_{F} q^{2}_{m}\over \pi } {d\omega \over \omega } \left ( 1 - 
{\omega \over E_{\nu }} + {1\over 2} {\left (\omega \over E_{\nu} 
\right )}^2 \right ) \quad \ln^3 \left ({2\omega \over q_{m}} \right ), 
\label{eq:8}
\end{equation}

\noindent where $\omega $ is the photon energy, $E_{\nu }$ is the initial 
neutrino energy, 
$q_{m}$ is the maximal momentum of the nucleus recoil. For the high
energy neutrinos, within the above approximation the total 
cross-section of the process is

\begin{equation}
\sigma \simeq {\left ({\alpha \over 2\pi } \right )}^3 \; {Z^{2}\over 27} 
\; {G^{2}_{F}q^{2}_{m}\over \pi } \; \ln^4 \left ({2E_{\nu }\over q_{m}} 
\right ).
\label{eq:9}
\end{equation}

\noindent For example, for a neutrino energy $E_{\nu }= 100$ GeV we have

\begin{equation}
\sigma \simeq Z^{2}~\cdot~1.6~\cdot~10^{-46} \; cm^{2}.
\label{eq:10}
\end{equation}

\noindent This small value of the cross-section makes it difficult to
observe the bremsstrahlung in the neutrino scattering by
the coulombian field of the nucleus. This is true even if one takes into
account the distinctive signature of the reaction as the
production of a high energy photon without any accompanying
particles. It must be noted that the same signature in the
neutrino reaction may correspond to the coherent production of
photons by nucleons of the nucleus~\cite{Rein1,Gersh}. However, the process we
consider has a narrower angular distribution of photons, $\theta 
< q_{m}/E_{\nu }$
instead of $\theta < \sqrt{q_m/E_{\nu}}$~\cite{Gersh,Rein2}. 
Moreover, it is necessary to
distinguish in the neutrino experiment between the electromagnetic
showers produced by photons and by recoiled electrons in the
process $\nu e \rightarrow \nu e$ which has a cross-section $10^{4}$ 
times larger than (10).

Nevertheless, we hope to overcome in the future the
experimental difficulties we have pointed out. Then the process
$\nu \gamma^{*} \rightarrow \nu \gamma $ we have discussed could 
be accessible to observation.
This process (one-loop at the minimum) could be one of the few tests
for the validity of higher-order perturbation theory in the
standard model of electroweak interaction.

\medskip

The authors are grateful to K.A.Ter-Martirosian for permanent
interest and stimulating discussions and to V.B.Svetovoy for
helpful remarks.

\newpage

%%%%%%%%%%%%%%%%%%%%%%%%%  Fig.1  %%%%%%%%%%%%%%%%%%%%%%%%%%%%%%%%%%%%%%%
\begin{figure}[ht]
\centerline{
\epsffile[122 534 267 646]{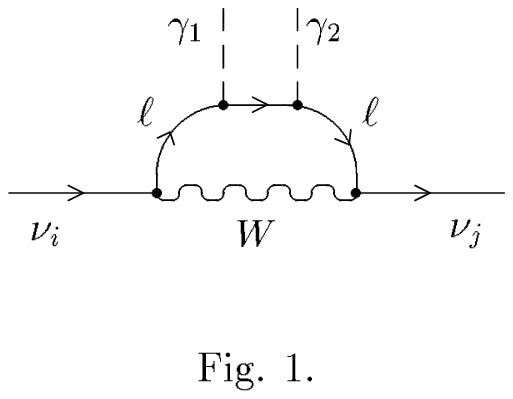}}
\end{figure}
%%%%%%%%%%%%%%%%%%%%%%%%%%%%%%%%%%%%%%%%%%%%%%%%%%%%%%%%%%%%%%%%%%%%%%%%%

\vspace{20mm}

%%%%%%%%%%%%%%%%%%%%%%%%%  Fig.2  %%%%%%%%%%%%%%%%%%%%%%%%%%%%%%%%%%%%%%%
\begin{figure}[ht]
\centerline{
\epsffile[163 521 251 637]{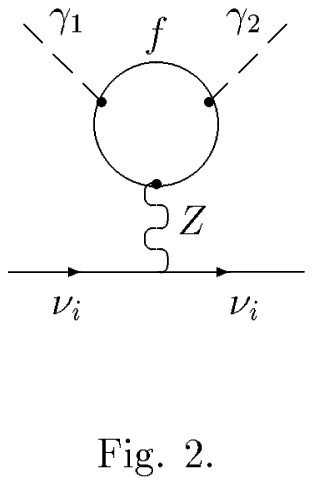}}
\end{figure}
%%%%%%%%%%%%%%%%%%%%%%%%%%%%%%%%%%%%%%%%%%%%%%%%%%%%%%%%%%%%%%%%%%%%%%%%%

\end{document}